\documentclass[showpacs,pra,twocolumn]{revtex4}
\usepackage{graphicx}
\usepackage{amsmath}
\usepackage{amsfonts}
\usepackage{amssymb}
\newcommand{\Real}{{\textrm{Re}}}

\newcommand{\Tr}{{\textrm{Tr}}} 

\newcommand{\ket}[1]{|#1\rangle}
\newcommand{\bra}[1]{\langle #1|}
\begin{document}
\title{Non-Abelian Generalization of Off-Diagonal Geometric Phases} 
\author{David Kult$^{1}$\footnote{Electronic address:
david.kult@kvac.uu.se}, Johan {\AA}berg$^{2}$\footnote{Electronic
address: J.Aberg@damtp.cam.ac.uk}, and Erik
Sj\"oqvist$^{1}$\footnote{Electronic address: eriks@kvac.uu.se}}
\affiliation{$^{1}$Department of Quantum Chemistry, Uppsala
University, Box 518, Se-751 20 Uppsala, Sweden. \\ $^{2}$Centre for
Quantum Computation, Department of Applied Mathematics and Theoretical
Physics, University of Cambridge, Wilberforce Road, Cambridge CB3 0WA,
United Kingdom.}
\date{\today}
\begin{abstract}
If a quantum system evolves in a noncyclic fashion 
the corresponding geometric phase or holonomy may not be fully
defined. Off-diagonal geometric phases have been developed to deal
with such cases. Here, we generalize these phases to the non-Abelian
case, by introducing off-diagonal holonomies that involve evolution of
more than one subspace of the underlying Hilbert space. Physical
realizations of the off-diagonal holonomies in adiabatic evolution and
interferometry are put forward.
\end{abstract}
\maketitle
\section{Introduction}
A quantal system that fails to return to its initial state after some
prescribed elapse of time may acquire a well-defined geometric phase
\cite{samuel88}. An interesting feature of this noncyclic geometric
phase is that it becomes undefined when the initial and final states
are orthogonal. This gives rise to a nodal point structure that can be
monitored experimentally in a history-dependent manner
\cite{bhandari91,bhandari92a,bhandari92b,bhandari93}. In the hope 
to recover some of the lost
interference information at the nodal points of the noncyclic
geometric phase, Manini and Pistolesi \cite{manini00} introduced
off-diagonal geometric phases for adiabatic evolutions of pure
states. These quantities may be defined in cases where the standard
geometric phase is not. The adiabatic requirement on the evolution in
ref. \cite{manini00} was lifted by Mukunda {\it et. al.}
\cite{mukunda02}, and Hasegawa {\it et. al.} \cite{hasegawa01,hasegawa02}
provided an experimental verification of the second order off-diagonal
geometric phase for neutron spin. Theories for
off-diagonal phases and holonomies for mixed quantal states have been
developed \cite{filipp03a,filipp03b,tong05,filipp05}.

Wilczek and Zee \cite{wilczek84} showed that the geometric phase
factor generalizes to a unitary state change, often referred to as a
non-Abelian quantum holonomy, when considering cyclic adiabatic
evolution governed by a degenerate Hamiltonian. The relevance of
non-Abelian holonomies for universal fault tolerant quantum
computation has been demonstrated in refs. 
\cite{zanardi99,pachos00a,pachos00b,duan01,recati02,faoro03}.  The
non-Abelian quantum holonomies have been generalized to nonadiabatic
\cite{anandan88}, discrete \cite{sjoqvist06}, and noncyclic evolutions
\cite{mostafazadeh99,kult06}. As for the geometric phase, the holonomy
may be undefined when the evolution is noncyclic. In the non-Abelian
case we also have the additional possibility that the holonomy is
partially defined \cite{kult06}.  

In this Letter, we extend ref. \cite{manini00} and introduce
non-Abelian off-diagonal holonomies. We demonstrate that the
off-diagonal holonomies retain holonomy information when the standard
noncyclic ones \cite{mostafazadeh99,kult06} are undefined. We also
suggest physical realizations of the off-diagonal holonomies using 
interferometry, both in adiabatic and nonadiabatic settings. 

\section{Off-diagonal Holonomies}
Consider a smoothly parameterized decomposition
\begin{eqnarray}
\mathcal{H} = \mathcal{H}_1(s) \oplus \cdots \oplus
\mathcal{H}_{\eta}(s), \ s \in [0,1] ,
\end{eqnarray}
of an $N$-dimensional Hilbert space $\mathcal{H}$ into $\eta$ mutually
orthogonal subspaces. It is to be noted that this parameter-dependent
decomposition can arise in different ways, e.g., it may consist of the
instantaneous eigenspaces of some $s$-dependent Hamiltonian
\cite{wilczek84}, or of some arbitrary decomposition that evolves
under the Schr\"odinger equation \cite{anandan88}. Assume that 
$\dim [\mathcal{H}_l(s)] = n_l$, $\forall s\in [0,1]$, and 
$l =1,\ldots,\eta$. Thus, each family $\mathcal{H}_l (s)$ of 
subspaces defines a curve $\mathcal{C}_l$ in the Grassmann manifold 
$\mathcal{G}(N;n_l)$, i.e.,
the set of $n_l$-dimensional subspaces in the $N$-dimensional Hilbert
space \cite{greub73}. For each such curve, we introduce the
quantities
\begin{equation}
\label{oldgammadef}
\Gamma_l = \lim_{\delta s\rightarrow 0}
P_l(1)P_l(1-\delta s)\cdots P_l(\delta s)P_l(0),
\end{equation}
where $P_l (s)$ is the projection operator onto the subspace 
$\mathcal{H}_l (s)$. This definition makes $\Gamma_{l}$ 
explicitly gauge invariant. Note also that the limit 
$\delta s\rightarrow 0$ in Eq.~(\ref{oldgammadef}) makes 
$\Gamma_{l}$ uniquely determined for any sufficiently smooth curve 
$\mathcal{C}_{l}$ in the Grassman manifold. We furthermore define 
\begin{eqnarray}
\label{sigmadef}
\sigma_{kl}= P_k (0)\Gamma_l .
\end{eqnarray}
Let $\{|k^i (s)\rangle\}_{i=1}^{n_k}$ and $\{|l^i
(s)\rangle\}_{i=1}^{n_l}$ be orthonormal bases for subspaces
$\mathcal{H}_k (s)$ and $\mathcal{H}_l (s)$, respectively, in terms of
which
\begin{eqnarray}
\label{sigmamatris}
\sigma_{kl} & = & \sum_{ij} [(\mathcal{F}^k_0|\mathcal{F}^l_1)
{\bf P} e^{\int_0^1 \boldsymbol{A}_l (s)ds}]_{ij} \ket{k^i (0)}
\bra{l^j (0)}
\nonumber \\
 & = & \sum_{ij} [\boldsymbol{\sigma}^{kl}]_{ij} \ket{k^i (0)}
\bra{l^j(0)} .
\end{eqnarray}
Here, $(\mathcal{F}^k_0|\mathcal{F}^l_1)$ is a $n_k \times n_l$ matrix
with components $(\mathcal{F}^k_0|\mathcal{F}^l_1)_{ij} = \bra{k^i
(0)} l^j (1) \rangle$ and $[\boldsymbol{A}_l(s)]_{ij} =
\bra{\partial_{s}l^i (s)} l^j (s) \rangle$ is the Wilczek-Zee
connection along $\mathcal{C}_l$ in $\mathcal{G} (N;n_l)$.

The unitary part $\Phi[\boldsymbol{\sigma}^{ll}]$\footnote{For a matrix 
$\boldsymbol{Z}$, $\Phi[\boldsymbol{Z}] \equiv (\sqrt{\boldsymbol{Z}
\boldsymbol{Z}^{\dagger}})^{\ominus}\boldsymbol{Z}$, $\ominus$ being
the Moore-Penrose pseudoinverse \cite{moore20,penrose55,lancaster85} 
obtained by inverting all nonzero eigenvalues of $\sqrt{\boldsymbol{Z}
\boldsymbol{Z}^{\dagger}}$. If $\det \sqrt{\boldsymbol{Z}
\boldsymbol{Z}^{\dagger}} \neq 0$ then the MP pseudoinverse of 
$\sqrt{\boldsymbol{Z} \boldsymbol{Z}^{\dagger}}$ coincides with 
the inverse.} is the holonomy in ref. \cite{kult06} associated 
with the (open) path $\mathcal{C}_l$.  It
seems natural to ask whether we can interpret the matrices
$\boldsymbol{\sigma}^{kl}, \ k\neq l,$ in a similar fashion. To answer
this, we need to see how these matrices behave under a gauge
transformation, i.e., a change of frames $\ket{l^i (s)} \rightarrow
| \widetilde{l}^i (s) \rangle = \sum_j | l^j(s)\rangle 
[\boldsymbol{U}_l (s)]_{ji}$, where 
$\{\boldsymbol{U}_l (s)\}_{l=1}^{\eta}$ are unitary matrices.  
Under such a transformation the matrices ${\bf P} 
e^{\int_0^1 \boldsymbol{A}_l(s)ds}$ and
$(\mathcal{F}^k_0|\mathcal{F}^l_1)$ undergo the following changes
\begin{eqnarray}
\label{transP}
{\bf P}e^{\int_0^1 \boldsymbol{A}_l (s)ds} & \rightarrow &
\boldsymbol{U}_l^{\dagger}(1){\bf P}e^{\int_0^1
\boldsymbol{A}_l(s)ds}\boldsymbol{U}_l (0),
\nonumber \\
(\mathcal{F}^k_0|\mathcal{F}^l_1) & \rightarrow &
\boldsymbol{U}_k^{\dagger}(0)
(\mathcal{F}^k_0|\mathcal{F}^l_1)\boldsymbol{U}_l(1) .
\end{eqnarray}
Consequently, $\boldsymbol{\sigma}^{kl}$ transforms as
\begin{eqnarray}
\label{transsigma}
\boldsymbol{\sigma}^{kl} \rightarrow
\boldsymbol{U}_k^{\dagger}(0) \boldsymbol{\sigma}^{kl}
\boldsymbol{U}_l(0) ,
\end{eqnarray}
i.e., noncovariantly unless $k=l$. Thus, the matrices
$\boldsymbol{\sigma}^{kl}$, $k\neq l$, fail to reflect the geometry of
the paths $\mathcal{C}_k$ and $\mathcal{C}_l$.  However, the specific
behavior of $\boldsymbol{\sigma}^{kl}$ under gauge transformations
suggests that we consider the operator
\begin{eqnarray}
\label{deforderkigen}
\gamma_{l_1 \ldots l_{\kappa}} &= & \sigma_{l_1 l_\kappa}
\sigma_{l_{\kappa} l_{\kappa-1}} \cdots
\sigma_{l_{3}l_{2}}\sigma_{l_2 l_1}
\nonumber \\
 & = & \sum_{ij} [\boldsymbol{ \gamma}^{l_1 \ldots l_{\kappa}}]_{ij}
\ket{l_1^i(0)} \bra{l_1^j(0)} ,
\end{eqnarray}
where $\boldsymbol{\gamma}^{l_1 \ldots l_{\kappa}}$ is the matrix
\begin{equation}
\label{deforderk2igen}
\boldsymbol{\gamma}^{l_1 \ldots l_{\kappa}} =
\boldsymbol{\sigma}^{l_1 l_\kappa}
\boldsymbol{\sigma}^{l_{\kappa}l_{\kappa-1}} \cdots
\boldsymbol{\sigma}^{l_{3}l_{2}}\boldsymbol{\sigma}^{l_{2} l_1}.
\end{equation}
We can use these operators and matrices to define gauge covariant
quantities, since $\Phi \big[ \boldsymbol{\gamma}^{l_1 \ldots
l_{\kappa}} \big] \rightarrow \boldsymbol{U}_{l_1}^{\dagger}(0) \Phi
\big[ \boldsymbol{\gamma}^{l_{1}l_{2}\ldots l_{\kappa}} \big]
\boldsymbol{U}_{l_1}(0)$ from eq. (\ref{transsigma}). Thus, we propose
to take
\begin{eqnarray}
\label{defhol}
\boldsymbol{U}_g^{(\kappa)} [\mathcal{C}_{l_1} , \ldots ,
\mathcal{C}_{l_{\kappa}}] =
\Phi \big[ \boldsymbol{\gamma}^{l_1 \ldots l_{\kappa}} \big]
\end{eqnarray}
as the gauge covariant non-Abelian holonomies of order $\kappa$,
and thus generalizing the approach of ref. \cite{manini00} to the
non-Abelian case. We extend the range of $\kappa$ by defining
$\boldsymbol{U}_g^{(1)} [\mathcal{C}_{l}] = \Phi \big[
\boldsymbol{\sigma}^{ll} \big]$, i.e., the first order ($\kappa =1$) 
holonomies are taken to be the open path holonomies in refs. 
\cite{mostafazadeh99,kult06}.

Note that the definition in eq. (\ref{defhol}) allows any sequence
$(l_1,\ldots,l_{\kappa})$. This includes cases like, e.g.,
$\boldsymbol{\gamma}^{111}$, which cannot be regarded as an
``off-diagonal" object. Hence, eq.~(\ref{defhol}) can be regarded as a
general definition of holonomies of degree $\kappa$, both diagonal and
off-diagonal. To define genuinely off-diagonal holonomies we obtain a
reasonable subclass if we require that $(l_{1},\ldots,l_{\kappa})$
contains each number at most once.  We let
$\mathbb{I}_{\kappa}^{\eta}$ denote all vectors
$(l_{1},\ldots,l_{\kappa})$ with $l_{j}\in\{1,\ldots,\eta\}$, such
that none of the numbers occurs twice, e.g., $(2,5,3)\in
\mathbb{I}_{3}^{6}$ but $(6,4,6,2)\notin \mathbb{I}_{4}^{6}$. We refer
to the set of holonomies $\boldsymbol{U}_g^{(\kappa)}
[\mathcal{C}_{l_{1}} , \ldots , \mathcal{C}_{l_{\kappa}}]$ with
$(l_{1},\ldots,l_{\kappa})\in \mathbb{I}_{\kappa}^{\eta}$ with $2\leq
\kappa\leq \eta$, as ``strictly off-diagonal holonomies".

For a cyclic evolution, characterized by $\mathcal{H}_l (1) =
\mathcal{H}_l (0), \ l =1,\ldots,\eta$, the standard holonomies
$\boldsymbol{U}_g^{(1)} [\mathcal{C}_l]$ are fully defined. On the
other hand, in this case we have
$\boldsymbol{\gamma}^{l_1 \ldots l_{\kappa}}=0$,
$(l_{1},\ldots,l_{\kappa})\in \mathbb{I}^{\eta}_{\kappa}$, $\kappa\geq
2$, which implies that all strictly off-diagonal holonomies are
undefined for cyclic evolution. Thus, just as in the Abelian case
\cite{manini00}, the standard holonomies contain all nontrivial
information about $\mathcal{C}_1,\ldots,\mathcal{C}_{\eta}$ when these
are loops.

In the case where $n_l = 1$, $l = 1,\ldots,\eta$, the matrices
$(\mathcal{F}^k_0|\mathcal{F}^l_1)$ and ${\bf P}e^{\int_0^1
\boldsymbol{A}_l (s)ds}$ reduce to the complex numbers
$\langle k (0)\ket{l (1)}$ and $e^{-\int_0^1 \langle l (s)
\ket{\partial_{s}l (s)}ds}$, respectively. This leads to the
off-diagonal geometric phase factors
\begin{eqnarray}
\label{1by1case}
 & & \boldsymbol{U}_g^{(\kappa)}
[\mathcal{C}_{l_{1}},\ldots,\mathcal{C}_{l_{\kappa}}] =
\Phi \big[ \langle l_1 (0) \ket{l_{\kappa} (1)}
\nonumber \\
 & &\times e^{-\int_0^1 \langle l_{\kappa} (s) |
\partial_{s}l_{\kappa} (s)\rangle ds} \cdots
\langle l_{2} (0) \ket{l_{1} (1)}
\nonumber \\
 & & \times
e^{-\int_0^1 \langle l_1 (s) \ket{\partial_{s}l_1 (s)}ds} \big] ,
\end{eqnarray}
which coincide with $\Phi \big[ \gamma^{(\kappa) \Gamma}_{l_1 l_{\kappa}
l_{\kappa-1} \ldots l_2} \big]$ in ref. \cite{manini00}.

Manini and Pistolesi \cite{manini00} suggest an interpretation of
their off-diagonal geometric phases in terms of Berry phases for
single closed paths. In the second order case, these paths consist of
the segments $\mathcal{C}_k$, $\mathcal{C}_l$, $G_{kl}$, and $G_{lk}$, 
where $G_{kl}$ geodesically connects the final point of
$\mathcal{C}_k$ with the starting point of $\mathcal{C}_l$, and vice
versa for $G_{lk}$ (see fig. 1 of ref.  \cite{manini00}).  In the
general non-Abelian case, however, this interpretation is difficult to
maintain. Apart from the special case when $n_1 = n_2 = \ldots =
n_{\kappa}$, it is not possible to join the curves
$\mathcal{C}_{l_1},\ldots,\mathcal{C}_{l_{\kappa}}$, due to the
mismatch of dimensions, and thus the closure using geodesics is not
applicable. This observation shows that the off-diagonal holonomies 
in general cannot be interpreted as standard Wilczek-Zee quantum 
holonomies for closed paths \cite{wilczek84}, and that the 
off-diagonal holonomies 
therefore are genuinely new concepts associated with the evolution 
of quantum systems. Another consequence of the fact that we can have 
different $n_l$ is that the rank of $\boldsymbol{\gamma}^{l_1 \ldots
l_{\kappa}}$ cannot be larger than the smallest $n_l$ (see 2.17.8 of
ref. \cite{marcus64}), and thus may be less than $n_{l_1}$.  

The non-Abelian character of the off-diagonal holonomies
$\boldsymbol{U}^{(\kappa)}
[\mathcal{C}_{1},\ldots,\mathcal{C}_{\kappa}]$ implies that they are
not invariant under cyclic permutations of the indexes
$(l_1, \ldots , l_{\kappa})$. It may even be the case that two
off-diagonal holonomies that differ only by a cyclic permutation have
different rank, since the smallest $n_l$ only provides an upper bound
for the rank of $\boldsymbol{\gamma}^{l_1 \ldots
l_{\kappa}}$. Furthermore, as is exemplified below it is also possible
that $\boldsymbol{\gamma}^{l_1 \ldots l_{\kappa}}$ may have 
path-dependent nodal points if $\kappa \geq 2$.

\section{Where did the phase information go?}
As noted above, all strictly off-diagonal holonomies are undefined for
cyclic evolutions, in analogy with the standard cyclic off-diagonal
geometric phases. Conversely, the noncyclic geometric phases are
undefined when the states at the initial and final points of the
curves are orthogonal, i.e., there exist nodal points where these 
phases are undefined. The hope to recover this lost phase
information appears to have been one of the primary reasons for Manini
and Pistolesi to introduce off-diagonal geometric phases. However, the
issue concerning the possible recovery of phase information was never 
explicitly investigated in ref. \cite{manini00}. In
the following we elucidate some aspects of this question in the case
of non-Abelian off-diagonal holonomies. 

Let us first analyze what happens if the rank of some of the overlap
matrices $(\mathcal{F}_0^l|\mathcal{F}_1^l)$ is greater than zero but
less than their subspace dimension $n_l$. This is a situation where
the corresponding holonomies become partial \cite{kult06}; a
phenomenon that has no counterpart in the Abelian case. To aid us in
this analysis we introduce the unitary $N\times N$ matrix
\begin{equation}
\label{totalsigma}
\boldsymbol{S}_{\textrm{tot}} =
\begin{pmatrix}
\boldsymbol{\sigma}^{11}  & \ldots &
\boldsymbol{\sigma}^{1\eta} \\
\vdots & \ddots & \vdots \\
\boldsymbol{\sigma}^{\eta 1} & \ldots &
\boldsymbol{\sigma}^{\eta \eta}
\end{pmatrix}.
\end{equation}
It follows from unitarity that $R(\boldsymbol{S}_{\textrm{tot}})=N$,
where $R(\boldsymbol{X})$ denotes the rank of matrix
$\boldsymbol{X}$. Furthermore, for every $l = 1, \ldots , \eta$,
it holds that $\sum_k
\boldsymbol{\sigma}^{kl\dagger}\boldsymbol{\sigma}^{kl} = \sum_k
\boldsymbol{\sigma}^{lk}\boldsymbol{\sigma}^{lk\dagger} =
\boldsymbol{1}_{n_l \times n_l}$, where $\boldsymbol{1}_{n_l \times
n_l}$ denotes the $n_l \times n_l$ identity matrix. This entails that
(see 2.17.2 and 2.17.5 of ref.  \cite{marcus64}) $\sum_k
R(\boldsymbol{\sigma}^{kl}) \geq n_l $ and $\sum_k
R(\boldsymbol{\sigma}^{lk}) \geq n_l $.  So, if
$R(\boldsymbol{\sigma}^{ll})= n_l - n$, then $\sum_{k\neq l}
R(\boldsymbol{\sigma}^{kl}) \geq n$ and $\sum_{k\neq l}
R(\boldsymbol{\sigma}^{lk}) \geq n$. In other words, when the overlap
matrix $(\mathcal{F}_0^l|\mathcal{F}_1^l)$ decreases by $n$ in rank,
the lower bound for the sum of the ranks of the matrices
$\boldsymbol{\sigma}^{kl}$ increases by the same amount. Thus, the
``holonomy information'' that is lost when the holonomy of the curve
$\mathcal{C}_l$ becomes partial is transferred to the matrices
$\boldsymbol{\sigma}^{kl}$.

A perhaps more significant question is whether the rank of the matrices
$\boldsymbol{\gamma}^{l_{1}\ldots l_{\kappa}}$ depends on the rank of
the overlap matrix $(\mathcal{F}_0^l | \mathcal{F}_1^l)$ in a
manner similar to what was discussed above for the matrices
$\boldsymbol{\sigma}^{kl}$. One can demonstrate with a simple
counterexample that no such relation exists. Assume
$\eta=3$ and $n_{1} = n_{2} = n_{3} = 2$. Furthermore, assume
that $\boldsymbol{\sigma}^{13}$, $\boldsymbol{\sigma}^{21}$, and
$\boldsymbol{\sigma}^{32}$ are the $2\times 2$ zero matrix, and
\begin{equation*}
{\boldsymbol{\sigma}^{11}}^{\dagger} =
{\boldsymbol{\sigma}^{22}}^{\dagger}={\boldsymbol{\sigma}^{33}}^{\dagger}
=\boldsymbol{\sigma}^{12} =
\boldsymbol{\sigma}^{23}=\boldsymbol{\sigma}^{31}=
\begin{pmatrix}
0 & 0 \\
1 & 0
\end{pmatrix}.
\end{equation*}
One may verify that the corresponding matrix
$\boldsymbol{S}_{\textrm{tot}}$ is unitary. In this example
$\boldsymbol{\gamma}^{kl} = 0 , \ \forall k,l$, and
$\boldsymbol{\gamma}^{klm}=0, \ \forall k,l,m$. Hence, although none
of the overlap matrices $(\mathcal{F}^l_0 | \mathcal{F}^l_1)$ are of
full rank, all strictly off-diagonal holonomies vanish, even those of
higher order.

Finally, we ask what happens if the matrices
$\boldsymbol{\sigma}^{ll}$ are zero for all $l =1,\ldots,\eta$. We
prove by {\it reductio ad absurdum} that at least one of the strictly
off-diagonal holonomies must have nonzero rank.  Assume
$\boldsymbol{\sigma}^{ll}=0$, for $l =1, \ldots, \eta$, and
$\boldsymbol{\gamma}^{l_1 \ldots l_{\kappa}}=0$, for $(l_{1}, \ldots
,l_{\kappa}) \in \mathbb{I}_{\kappa}^{\eta}$, $\kappa\geq 2$. Consider
an arbitrary string $(l_{1}, \ldots, l_{\nu})$ consisting of the
integers $1$ to $\eta$, and with $\nu\geq 2$.  If $(l_{1}, \ldots,
l_{\nu})\in \mathbb{I}_{\nu}^{\eta}$ ($\nu \leq\eta$), then by
assumption $\boldsymbol{\gamma}^{l_{1} \ldots l_{\nu}}=0$.  If
$(l_{1}, \ldots, l_{\nu}) \notin \mathbb{I}_{\nu}^{\eta}$, then take
one of the smallest subsequences $(l_{\alpha},l_{\alpha+1},\ldots,
l_{\beta-1}, l_{\beta})$ that begins and ends with the same number
(i.e., $l_{\beta} = l_{\alpha}$). In this subsequence there is no
other repetitions (otherwise there exists a smaller subsequence). It
follows that $\boldsymbol{\gamma}^{l_{1} \ldots l_{\nu}} =
\boldsymbol{\sigma}^{l_{1}l_{\nu}}\cdots
\boldsymbol{\sigma}^{l_{\beta+1}l_{\beta}}
\boldsymbol{\gamma}^{l_{\alpha}\ldots l_{\beta-1}}
\boldsymbol{\sigma}^{l_{\alpha}l_{\alpha-1}} \cdots
\boldsymbol{\sigma}^{l_{2}l_{1}}$. Moreover, $(l_{\alpha}, \ldots ,
l_{\beta-1}) \in \mathbb{I}^{\eta}_{\beta-\alpha-1}$. If $\beta-\alpha
= 1$, then $\boldsymbol{\gamma}^{l_{\alpha} \ldots l_{\beta-1}} =
\boldsymbol{\sigma}^{l_{\alpha}l_{\alpha}} =0$, otherwise
$\boldsymbol{\gamma}^{l_{\alpha}\ldots l_{\beta-1}} =0$ by
assumption. We proceed by noting that $\Tr
\left(\boldsymbol{S}_{\textrm{tot}}^{\nu} \right)= \sum_{l_{1} \ldots
l_{\nu}} \Tr \boldsymbol{\gamma}^{l_{1} \ldots l_{\nu}} = 0$, for all
$\nu= 1,2,\ldots$, as a consequence of our assumptions.  However, this
cannot be the case since $\boldsymbol{S}_{\textrm{tot}}$ is a unitary
matrix. Therefore, our assumptions must be wrong, and at least one of
the strictly off-diagonal holonomies must have nonzero rank. By
the above findings we can conclude that the holonomy information lost
in the the nodal points of the non-Abelian noncyclic holonomies
indeed can be retained in some sense, but that the structure is much
more intricate and rich than in the Abelian case, due to the 
existence of partial holonomies in the non-Abelian setting.

\section{Example}
We illustrate the off-diagonal holonomies by an example in the
adiabatic context.  We let the system evolve under the action of a
slowly varying Hamiltonian. Let us consider the tripod system
\cite{duan01,unanyan99} modeled by the parameter-dependent four-state
Hamiltonian $H(s) = \omega
\ket{e} (\sin \theta (s) \cos \varphi (s) \bra{0} +
\sin \theta (s) \sin \varphi (s) \bra{1} + \cos \theta (s)\bra{a}) +
\textrm{h.c.}$, exhibiting two nondegenerate `bright' states
$|\mathcal{F}_s^{\pm} ) = \{ \ket{B^{\pm} (s)} \}$ with energy
$\pm \omega$ and a doubly degenerate `dark' zero energy eigenspace
$|\mathcal{F}_s^d ) = \{ \ket{D^1 (s)} , \ket{D^2 (s)} \}$.
Explicitly, we may choose
\begin{eqnarray}
\ket{B^{\pm}} & = & \frac{1}{\sqrt{2}} \Big( \ket{e} \pm
\sin \theta \cos \varphi \ket{0}
\pm \sin \theta \sin \varphi \ket{1}
\nonumber \\
 & & \pm \cos \theta \ket{a} \Big) ,
\nonumber \\
\ket{D_1} & = & \cos \theta \cos \varphi \ket{0} +
\cos \theta \sin \varphi \ket{1} - \sin \theta \ket{a} ,
\nonumber \\
\ket{D_2} & = & - \sin \varphi \ket{0} + \cos \varphi \ket{1} .
\end{eqnarray}
Consider paths $(0,0) \rightarrow (\theta_1,\varphi_1)$ in parameter
space $(\theta,\varphi)$. For each such path the energy eigenstates
define paths $\mathcal{C}_{\pm}$ in $\mathcal{G} (4;1)$ and
$\mathcal{C}_d$ in $\mathcal{G} (4;2)$. We obtain the geometric phase
factors $\boldsymbol{U}^{(1)} [\mathcal{C}_{\pm}] = 1$ for $\theta_1
\neq \pi$ and $\boldsymbol{U}^{(2)} [\mathcal{C}_{\pm} ,
\mathcal{C}_{\mp}] = 1$ for $\theta_1 \neq 0$.  $\boldsymbol{U}^{(1)}
[\mathcal{C}_{\pm}]$ are undefined at $\theta_1 = \pi$ and similarly
$\boldsymbol{U}^{(2)} [\mathcal{C}_{\pm} ,\mathcal{C}_{\mp}]$ at
$\theta_1 =0$.  In ref. \cite{kult06} it was shown that
$\boldsymbol{U}^{(1)} [\mathcal{C}_d]$ is fully defined, except when
the path ends at $\theta_1 = \pi/2$, where the holonomy becomes
partial. The strictly off-diagonal holonomies ($\kappa = 2,3$)
involving the dark subspace are undefined when $\sin \theta_1 = 0$.
For $\sin \theta_1 \neq 0$, let $Z
= \int_0^1
\cos [\theta (s)] \dot{\varphi} (s) ds$ and we obtain
\begin{eqnarray}
\boldsymbol{U}^{(2)} [\mathcal{C}_{\pm}, \mathcal{C}_d] & = &
- \boldsymbol{U}^{(3)} [\mathcal{C}_{\pm},
\mathcal{C}_{\mp}, \mathcal{C}_d]
\nonumber \\
 & = & - \boldsymbol{U}^{(3)}
[\mathcal{C}_{\pm}, \mathcal{C}_d, \mathcal{C}_{\mp}] =
- \ \frac{\cos (\varphi_1 - Z)}{|\cos (\varphi_1 - Z)|} ,
\nonumber \\
\boldsymbol{U}^{(2)} [\mathcal{C}_d, \mathcal{C}_{\pm}] & = &
- \boldsymbol{U}^{(3)} [\mathcal{C}_d, \mathcal{C}_{\pm},
\mathcal{C}_{\mp}]
\nonumber \\
 & = & -
\begin{pmatrix}
\cos Z \cos \varphi_1 &
\sin Z \cos \varphi_1 \\
\cos Z \sin \varphi_1 &
\sin Z \sin \varphi_1
\end{pmatrix} .
\end{eqnarray}
While $\boldsymbol{U}^{(2)} [\mathcal{C}_d, \mathcal{C}_{\pm}]$
and $\boldsymbol{U}^{(3)} [\mathcal{C}_d, \mathcal{C}_{\pm},
\mathcal{C}_{\mp}]$ are nonzero partial isometries, there are 
path-dependent nodal points of
$\boldsymbol{U}^{(2)} [\mathcal{C}_{\pm}, \mathcal{C}_d]$,
$\boldsymbol{U}^{(3)} [\mathcal{C}_{\pm},
\mathcal{C}_{\mp}, \mathcal{C}_d]$, and $\boldsymbol{U}^{(3)}
[\mathcal{C}_{\pm}, \mathcal{C}_d, \mathcal{C}_{\mp}]$,
namely where $\cos (\varphi_1 - Z) = 0$.

\section{Physical realizations}

\begin{figure}
\includegraphics[width = 8.5cm]{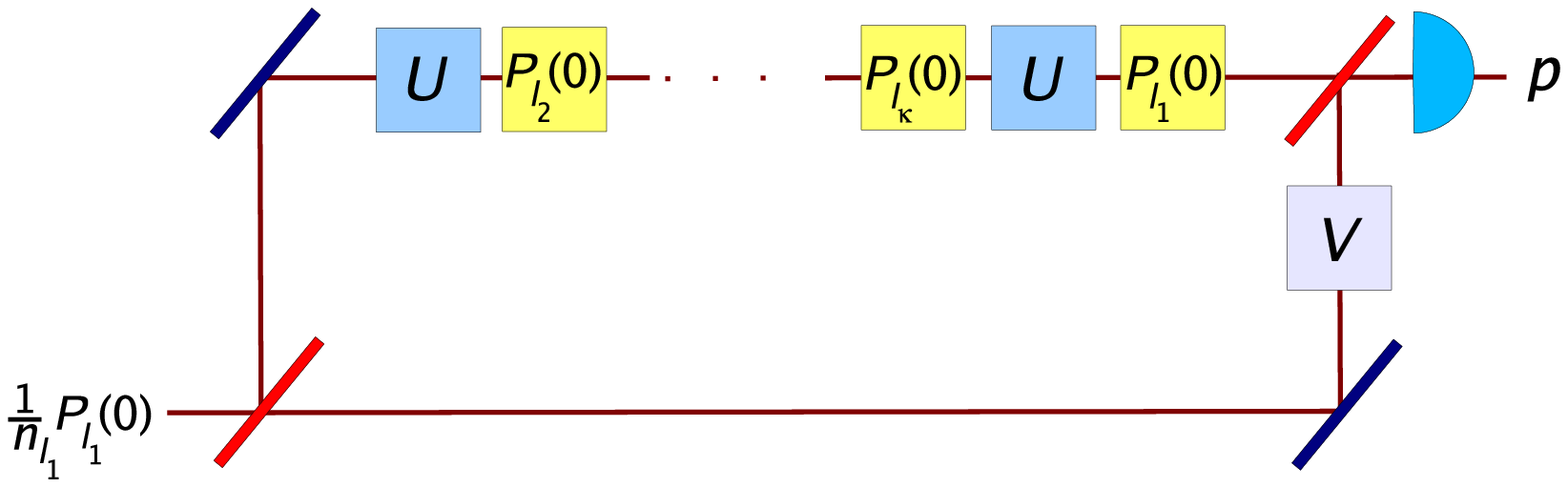}
\caption{\label{fig:fig1} Interferometric approach to
obtain the holonomy $\boldsymbol{U}_g^{(\kappa)} [\mathcal{C}_{l_1},
\ldots, \mathcal{C}_{l_{\kappa}}]$. The horizontal and vertical paths
of the Mach-Zehnder interferometer correspond to the states
$|0\rangle$ and $|1\rangle$, respectively. The particle have an
internal degree of freedom, on which the unitary operation $U$ is
applied. $U$ can be generated either by adiabatic or nonadiabatic
evolution, or approximated by filtering measurements. The particle
enters the interferometer in path $0$ and in the internal state
$P_{l_{1}}(0)/n_{l_{1}}$. In path $1$ a variable unitary operator $V$
such that $[V,P_{l}(0)]=0$, $\forall l$, is applied,  while in path
$0$ we alternatingly apply $U$ and a filtering corresponding to the
projector $|0\rangle\langle 0|\otimes P_{l_k} (0) + |
1\rangle\langle1|\otimes\hat{1}$, for $k=2,\ldots \kappa$. A final $U$
is applied, followed by the filtering $|0\rangle\langle 0|\otimes
P_{l_1} (0) + |1\rangle\langle1|\otimes\hat{1}$. The probability $p$
to find the particle in path $0$ is measured after a second beam
splitter. The unitary operator $V$ is varied so as to maximize $p$.}
\end{figure}

Let us now examine some possible physical realizations of
$\boldsymbol{U}_g^{(\kappa)} [\mathcal{C}_{l_{1}} , \ldots,
\mathcal{C}_{l_{\kappa}}]$. Consider the Mach-Zehnder interferometer
in fig. \ref{fig:fig1} with the two path states represented by
$\ket{0}$ and $\ket{1}$. We let the internal state of the particle
(e.g., spin) be represented by the Hilbert space
$\mathcal{H}=\mathcal{H}_1(s) \oplus \cdots \oplus
\mathcal{H}_{\eta}(s)$, $s \in [0,1]$.  The total system is prepared
in the state $|0\rangle\langle 0|\otimes P_{l_{1}}(0)/n_{l_{1}}$. We
first apply a beam-splitter, followed by the unitary operations
$\ket{0} \bra{0} \otimes \hat{1} + \ket{1} \bra{1} \otimes V$ and
$\ket{0} \bra{0} \otimes U + \ket{1} \bra{1} \otimes \hat{1}$, where
$V$ is a variable unitary operator assumed to be chosen such that
$[V,P_l(0)]=0$ for all $l$. Next, we perform a filtering corresponding
to the projection operator $\ket{0} \bra{0} \otimes P_{l_{2}}(0) +
\ket{1} \bra{1} \otimes \hat{1}$, i.e., the particle is ``removed" if
it is found in path $0$ with its internal state outside subspace
$\mathcal{H}_{l_{2}}$. Thereafter, we again apply the operator
$\ket{0} \bra{0} \otimes U + \ket{1} \bra{1} \otimes \hat{1}$, and the
filtering $\ket{0} \bra{0} \otimes P_{l_{3}}(0) + \ket{1} \bra{1}
\otimes \hat{1}$. This procedure is repeated until we have applied the
operator $\ket{0} \bra{0} \otimes U + \ket{1} \bra{1} \otimes
\hat{1}$, $\kappa$ times. After this, we apply a final filtering
$\ket{0} \bra{0} \otimes P_{l_{1}}(0) + \ket{1} \bra{1} \otimes
\hat{1}$, and recombine the two paths with a beam splitter. We
finally measure the probability $p$ to find the particle in path $0$.

First, we assume that the unitary operator $U$ acting on
$\mathcal{H}$ is caused by an adiabatic evolution of a time-dependent
Hamiltonian with eigenspaces $\{\mathcal{H}_l(s)\}_{l=1}^{\eta}$.  This
allows us to write $U = \sum_{l}e^{i\phi_{l}}\Gamma_{l}$, where
$\phi_l$ is the dynamical phase $\phi_l = \int_{0}^{1}E_{l}(s)ds$, and
$E_{l}(s)$ the eigenvalue corresponding to eigenspace
$\mathcal{H}_{l}(s)$ of the Hamiltonian. The corresponding
detection probability becomes
\begin{eqnarray}
\label{proba}
p & = & \frac{1}{4} + \frac{1}{4}
\frac{1}{n_{l_{1}}}\Tr(\gamma_{l_{1}\ldots
l_{\kappa}}\gamma_{l_{1}\ldots l_{\kappa}}^{\dagger})
\nonumber \\
 & & + \frac{1}{2}
\frac{1}{n_{l_{1}}}\Real[e^{i\sum_{k=1}^{\kappa}\phi_{k}}
\Tr(\boldsymbol{\gamma}^{l_{1}\ldots
l_{\kappa}}\boldsymbol{V}^{\dagger})],
\end{eqnarray}
where $\boldsymbol{V}_{ij} = \langle l_1^i (0)|V| l_1^j
(0)\rangle$. Note that $\boldsymbol{V}$ is a unitary matrix since
$[V,P_{l_1}(0)]=0$. By varying $V$ we obtain the
maximal detection probability when $\boldsymbol{V} =
e^{i\sum_{k=1}^{\kappa}\phi_{k}}\boldsymbol{U}_g^{(\kappa)}
[\mathcal{C}_{l_{1}} , \ldots, \mathcal{C}_{l_{\kappa}}]$. Hence, up
to the dynamical phases we have found the holonomy. 

In the adiabatic setting there is in the general case no easy way to
eliminate the dynamical phases. To avoid these problems we consider
two alternative approaches to generate the unitary operator $U$. One
alternative is to base the evolution entirely on filtering, where we
approximate the evolution in the spirit of ref. \cite{sjoqvist06}.
We begin with the same initial state, beam-splitter, and variable
unitary $V$, as in the previous case. Next we apply a sequence of
filterings $\ket{0} \bra{0} \otimes P_{l_{1}}(s_{j}) + \ket{1} \bra{1}
\otimes \hat{1}$, where $s_{j}$ form a discretization of the interval
$[0,1]$. For the next step we apply the sequence of filterings
$\ket{0} \bra{0} \otimes P_{l_{2}}(s_{j}) + \ket{1} \bra{1} \otimes
\hat{1}$, and we continue up $l_{\kappa}$. We finally apply $\ket{0}
\bra{0} \otimes P_{l_{1}}(0) + \ket{1} \bra{1} \otimes \hat{1}$,
followed by a beam splitter, and measure the probability to find the
particle in path $0$. One can show that the probability is
\begin{eqnarray}
\label{proba2}
p & = & \frac{1}{4} + \frac{1}{4}
\frac{1}{n_{l_{1}}}\Tr(\gamma_{l_{1}\ldots
l_{\kappa}}\gamma_{l_{1}\ldots l_{\kappa}}^{\dagger})
\nonumber \\
 & & + \frac{1}{2}
\frac{1}{n_{l_{1}}}\Real[\Tr(\boldsymbol{\gamma}^{l_{1}\ldots
l_{\kappa}}\boldsymbol{V}^{\dagger})].
\end{eqnarray}
Hence,  as in eq. (\ref{proba}),
apart from the absence of dynamical phases.

The second alternative that allows us to avoid the problem with
dynamical phases is to use a nonadiabatic approach.  Assume that the
evolution is driven by the time-dependent Hamiltonian $H(s)$, where
now $s$ is the time-parameter. We let the subspaces
$\mathcal{H}_{l}(s)$ be evolving under $H(s)$ according to the
Schr\"odinger equation. In contrast to the adiabatic approach the
subspaces $\mathcal{H}_{l}(s)$ are in the general case not eigenspaces
of $H(s)$. We furthermore let $\{|l^{j}(s)\rangle\}_{j=1}^{n_{l}}$ be
smoothly parameterized orthonormal bases of the subspaces
$\mathcal{H}_{l}(s)$. We wish to find the unitary matrices
$\boldsymbol{U}_l(s)$ such that the vectors
\begin{eqnarray}
|\chi_{l}^{k}(s)\rangle =
\sum_{j}|l^{j}(s)\rangle [\boldsymbol{U}_l(s)]_{jk}
\label{eq:vectors}
\end{eqnarray} 
satisfy the Schr\"odinger equation $i\partial_{s}|\chi_{l}^{k}(s)\rangle =
H(s)|\chi_{l}^{k}(s)\rangle$ ($\hbar =1$) with initial conditions
$|\chi_{l}^{k}(0)\rangle = |l^{k}(0)\rangle$.  If we substitute 
eq. (\ref{eq:vectors}) into
the Schr\"odinger equation we find that $\boldsymbol{U}_l(s)$ has to
satisfy $i\partial_{s}\boldsymbol{U}_l(s) =
i\boldsymbol{A}_{l}(s)\boldsymbol{U}_l(s) +
\boldsymbol{K}_{l}(s)\boldsymbol{U}_l(s)$, where
$[\boldsymbol{A}_{l}(s)]_{j'j} = \langle
\partial_{s}l^{j'}(s)|l^{j}(s)\rangle$ and 
$[\boldsymbol{K}_{l}(s)]_{j'j} =\langle
l^{j'}(s)|H(s)|l^{j}(s)\rangle$ contain the geometrical 
and dynamical contributions, respectively, 
as was discussed in ref. \cite{anandan88}. In order to get rid of the
dynamical contribution without affecting the evolution of the
subspaces we introduce the modified time-dependent Hamiltonian
\begin{equation}
\overline{H}(s)= H(s)- \sum_{l=1}^{\eta} P_{l}(s)H(s)P_{l}(s). 
\end{equation}
The evolution of the subspaces
$\mathcal{H}_{l}(s)$ are not affected by this modification since
$[\overline{H}(s),P_l(s)]=[H(s),P_l(s)]$, $\forall l$. We now wish to
find the unitary matrices $\overline{\boldsymbol{U}}_l(s)$ such that
the vectors $|\overline{\chi}^{k}_l(s)\rangle =
\sum_{j}|l^{j}(s)\rangle [\overline{\boldsymbol{U}}_l(s)]_{jk}$
satisfy the modified Schr\"odinger equation
$i\partial_{t}|\overline{\chi}^{k}_l(s)\rangle = 
\overline{H}(s)|\overline{\chi}^{k}_l(s)\rangle=[H(s)-
P_{l}(s)H(s)P_{l}(s)]|\overline{\chi}^{k}_l(s)\rangle$ with initial
conditions $|\overline{\chi}^{k}_l(0)\rangle = |l^{k}(0)\rangle$. In
this case we obtain $\partial_{s}\overline{\boldsymbol{U}}_l(s) =
i\boldsymbol{A}_{l}(s)\overline{\boldsymbol{U}}_l(s)$. Hence, the
solution $\overline{\boldsymbol{U}}_l(s)= {\bf
P}e^{\int_{0}^{s}\boldsymbol{A}_{l}(s')ds'}$ only depends on the geometric
contribution\footnote{One may note, though, that 
$\overline{\boldsymbol{U}}_l(s)$ is not gauge covariant and can 
therefore not be considered a geometric quantity, see ref. 
\cite{kult06}.}. 
The time-dependent Hamiltonian $\overline{H}(s)$ generates a unitary 
mapping from the initial state to the state at time $s =1$ that is 
given by $\overline{U}(1)=\sum_l \Gamma_l$, where
\begin{eqnarray}
\Gamma_{l} &=& \sum_{k}|\overline{\chi}^{k}_l(1)\rangle\langle
\overline{\chi}^{k}_l(0)| \nonumber \\ &=& \sum_{jk}[{\bf
P}e^{\int_{0}^{1}\boldsymbol{A}(s)ds}]_{jk}|l^{j}(1)\rangle \langle
l^{k}(0)|.
\end{eqnarray}
This means that if we let $U=\overline{U}(1)$ in the alternating 
procedure described above (see fig. \ref{fig:fig1}), then the 
probability to detect the particle in path $0$ becomes as in eq. 
(\ref{proba2}), and is maximized when $\boldsymbol{V} =
\boldsymbol{U}_g^{(\kappa)}[\mathcal{C}_{l_{1}} , \ldots,
\mathcal{C}_{l_{\kappa}}]$.
 
\section{Conclusion}
Noncyclic evolution of quantum systems may lead to
well-defined off-diagonal holonomies that involve more than one
subspace of Hilbert space. These holonomies reduce to the off-diagonal
geometric phases in ref. \cite{manini00} for one-dimensional
subspaces. The off-diagonal holonomies are undefined for cyclic
evolution but must contain members of nonzero rank when all the
standard holonomies are undefined. While the nodal point structure of
the holonomy for an open continuous path \cite{kult06} can only depend
on the end-points of the path, this structure can be path-dependent in
the off-diagonal case. Furthermore, we have put forward physical
realizations of the off-diagonal holonomies in the context of
adiabatic evolution and interferometry that may open up the
possibility to test these quantities experimentally.

\acknowledgments
J.{\AA}. thanks the Swedish Research Council for financial support and
the Centre for Quantum Computation at DAMTP, Cambridge, for
hospitality. E.S. acknowledges financial support from the Swedish
Research Council.

\end{document}